\documentclass[twocolumn,showpacs]{revtex4}
 \usepackage{epsfig,amsmath,amsfonts,psfrag,rotating}
 \newcommand{\nts}[1]{\tmspace{-}{#1\thinmuskip}{#1\txtmu}}
  \newcommand\eqlabel[1]{\label{#1}}

\allowdisplaybreaks
\begin{document}

\title{Beyond the random phase approximation in the Singwi-Sj\"olander theory of the half-filled Landau level}

\author{J. Dietel and W.Weller}
\affiliation{
Institut f\"ur Theoretische Physik, Universit\"at Leipzig,\\
Augustusplatz 10, D 04109 Leipzig, Germany \\
e-mail: dietel@itp.uni-leipzig.de\\}

\begin{abstract}
We study the $ \nu=1/2 $ Chern-Simons system and consider
a self-consistent field theory of the Singwi-Sj\"olander type
which goes beyond the random phase approximation (RPA). 
By considering the Heisenberg equation of motion for the longitudinal momentum
operator, we are able to show that the zero-frequency density-density response
function vanishes linearly in long wavelength limit independent
of any approximation. From this analysis, we
derive a consistency condition for a decoupling of the equal time
density-density and
density-momentum correlation functions. By using the
Heisenberg equation of motion of the Wigner distribution function
with a decoupling of the correlation functions which
respects this consistency condition, we calculate the response functions of the
$ \nu=1/2 $ system. In our scheme, we get a density-density response function which 
vanishes linearly in the Coulomb case for zero-frequency
in the long wavelength limit. Furthermore, we derive the compressibility,
and the
Landau energy as well as the Coulomb energy. These energies are in better
agreement with numerical and exact results, respectively,
than the energies calculated in the RPA.
\end{abstract}


\maketitle

\section{Introduction}
The combination of an electronic interaction and a strong magnetic field
in a two--dimensional electron system yields a rich variety of phases.
These are best classified by the filling factor $\nu$, which is the electron
density divided by the density of a completely filled Landau level.
In the case of $\nu \cong 1/2$, the behavior of the system resembles that of
a Fermi liquid in the absence of a magnetic field, or at small magnetic
fields. This effect can be explained with a new sort of quasi-particles:
at $\nu =1/2$, each electron combines with two flux quanta of the magnetic
field to form a composite fermion; these composite fermions then move
in an effective magnetic field which is zero on the average.
The interpretation of many experiments supports this picture.
We mention transport experiments with anti-- dots,
in which features of the resistivity are related to
closed loops of the composite fermions around the dots \cite{kan1},
and also focusing experiments \cite{sme1}. An overview over further
experiments can be found in \cite{wil1}. 
A field theoretical formulation of this composite fermion  picture was first
established by 
Halperin, Lee, and Read (HLR) \cite{hlr} as well as Kalmeyer and Zhang
\cite{ka1}.
They formulated the Hamiltonian in terms of Chern--Simons (CS)
transformed electrons and studied within the random--phase
approximation (RPA) many physical quantities.
Besides the theory of HLR there are
other alternative formulations of the composite fermionic picture
which are mainly based on a gauge transformation of the CS
Hamiltonian \cite{sh1}.

We used in \cite{di1} the theory of HLR for a determination of the ground
state energy of the $  \nu=1/2 $ system in RPA. There we found an infrared
diverging Landau energy. This problem was solved in \cite{ap1,di2,ko1}
by taking into account the correct normal ordering of the
operators. We obtained a Landau as well as a Coulomb energy in RPA
which is in satisfactory agreement with the exact and 
numerical results, respectively \cite{di2,ko1}. In \cite{co1}, Conti and Chakraborty (CC)
improved the calculation of the Coulomb energy by taking into
account dynamical correlations through the formalism
of Singwi, Tosi, Land and  Sj\"olander \cite{si1} (STLS),
known as the Singwi-Sj\"olander theory and established first in the
calculation of
the structure factor of the Coulomb theory. This method is a
generalization of
the RPA. 
In comparison to the RPA this method 
results in a Coulomb energy which is a better approximation
to the Coulomb energy
obtained earlier by numerical simulations of interacting electrons in the
lowest Landau level by Morf and d'Ambrumenil \cite{mo1} and by Girlich
\cite{gi1}.
CC did not calculate the Landau
energy of the $ \nu=1/2 $ system which would also be a very interesting
quantity  to be
compared with the RPA as well as the true result 
$ \omega_c/2 $, the Landau energy per electron.
In their theory the resulting zero-frequency
density-density response function vanishes as
the square of the wave vector in the long wavelength limit. This is in
contradiction to the  RPA result where it vanishes linearly for the Coulomb
interaction \cite{hlr}. In their paper, CC mentioned that the behaviour of the
zero-frequency response function of their theory
is similar to the zero-frequency response
functions of the alternative formulations of the CS theory \cite{sh1}
cited above.
But later on,
Halperin et al. \cite{st2} showed that the zero-frequency response
function of these theories also vanishes linearly in the wave vector
for the Coulomb
interaction if one expands the approximation to the RPA.

In this paper, we will show that the quadratic behaviour of the zero-frequency
response function in the theory of CC results from the  
decoupling of
the equal time  density-density and density-momentum correlation function. 
In CS-theories, this  needs a careful treatment
as will be shown by considering  the Heisenberg equation of motion for the
longitudinal momentum operator. With the help of this equation, we will get
a zero-frequency response function of the $ \nu=1/2 $ system 
which vanishes linearly for the Coulomb interaction in the
long-wavelength limit.
It can be also learned from this procedure 
which relation in the decoupling approximations of density-density
and density-momentum correlation functions
in CS theories should be fulfilled.
In the STLS theory it is the Heisenberg equation of motion of the Wigner
distribution function which is
used to get the response functions of the system. In CS-theories,
one has to decouple simultaneously a density-density correlation function 
as well as a density-momentum correlation function.
We will show that the consistency relation for the decoupling of these two
functions are not fulfilled in the theory of CC.
This is the reason for the quadratic behaviour of the long wavelength
zero-frequency density-density response function in their theory. 
We will suggest a STLS type theory of the CS system, in which the decouplings
respect the
consistency condition. With the help of this decoupling, we
will calculate the density-density  response function, the compressibility, the dynamic structure factor,
the static structure factor and the Coulomb energy,
as well as the Landau energy.

The paper is organized as follows:
We will introduce in section II the CS-Hamiltonian and
calculate the Heisenberg equation of motion for the longitudinal momentum.
Then we get from this equation the static density-density response
function in the long wavelength limit, and discuss the consistency 
relation for a decoupling of the equal time density-density and
density-momentum
correlation functions. We will study in section III
the equation of motion of the
Wigner distribution function and suggest a decoupling 
which respects this relation.

\section{Static response function of the
  $ \nu=1/2 \; $ CS system}

At first, we review the main steps of the HLR approach of the $ \nu=1/2 $
system. Here, we follow in the main the notation of CC \cite{co1}.
The CS transformation for spinless fermions is defined by \cite{hlr}
\begin{equation}  \eqlabel{10}
 \Psi^+(\vec{r})=\Psi^+_{e}(\vec{r})
 \exp\left[2i\int d\vec{r}\,' \mbox{arg}(\vec{r}-\vec{r}\,') \,
 \rho(\vec{r}\,')\right] \,.   
\end{equation}
where $\Psi^+_{e}(\vec{r}) $ is the electron creation operator,
$\Psi^+(\vec{r})$ is the transformed fermions,
$ \rho(\vec{r}) $ is the density operator of the fermion
operators , and
$ \mbox{arg}(\vec{r}) $  is the angle that $ \vec{r} $ forms with the
$ x $-axis. The kinetic part of the Hamiltonian is given after the
transformation as
\begin{equation}  \eqlabel{20}
  H_{\mbox{\scriptsize  kin}}=\frac{1}{2m_b}\int d^2r \, \Psi^+(\vec{r})
\left(-i \vec{\nabla}+\delta \vec{A}(\vec{r})\right)^2
\Psi(\vec{r}) \,.  
\end{equation}
where $ m_b $ is the electron band mass and
\begin{equation}      \eqlabel{30}
  \delta A_i(\vec{r})=\int d \vec{r}\,' \phi_i(\vec{r}-\vec{r}\,')
 \,  (\rho(\vec{r}\,')-\rho_0)\,.   
\end{equation}  
is the fluctuation of the CS vector potential.
$ \rho_0 $ is the mean density of the $ \nu=1/2 $ system and 
$ \vec{\phi}(\vec{r})=2 \, \vec{\nabla} \mbox{arg}(\vec{r})=
  2 \, \vec{e}_z \times \vec{r}/r^2 $. We used the convention $ \hbar=1 $
  and $ c=1 $ in the above formula (\ref{20}).
By expanding the Hamiltonian in (\ref{20}) and keeping only terms up to 
second order in the density fluctuations, one gets in the momentum space 
\begin{align}
&  H = \sum\limits_{\vec{k}} \frac{k^2}{2m_b}
 a^+_{\vec{k}} a_{\vec{k}}
 +\sum\limits_{\vec{k}\not=0} i \frac{v_1(k)}{m_b}
\nts{1}  \nts{1} \Bigg[:\left(\frac{\vec{k}}{k} \times \vec{\pi}(\vec{k})\right)
\rho(-\vec{k}) \nts{1} :\nts{1} \nonumber \\
&  \qquad +
\frac{1}{2} \left(v_0(k)+v_2(k)\right) \nts{1}  :\nts{1} \rho(\vec{k})
 \rho(-\vec{k}) \nts{1}:\nts{1} \Bigg] \,. \eqlabel{40}
\end{align}
Here $ \vec{\pi}(\vec{k}) $ is the Fourier transformed momentum operator
$ \vec{\pi}(\vec{r})=-i \Psi(\vec{r})^+  \vec{\nabla} \Psi(\vec{r}) $,
$ a^+_{\vec{k}} $ creates a CS-fermion with momentum $ \vec{k}$ 
and $ v_0(k)=2 \pi \epsilon^2/k $ is the Coulomb interaction
where $ \epsilon^2=e^2 /\varepsilon $.
$ e $ is the charge of the electron and
$ \varepsilon $ is the dielectric constant of the background. 
$ v_1(k)=4 \pi/k $ and
$ v_2(k)=(4\pi)^2 \rho_0/(m_b k^2)$ are CS potentials. We denote by $:A : $
the normal ordering of the operator $ A $. 
With the help of
the CS transformation of the electronic Hamiltonian, we thus get 
a CS Hamiltonian which does not contain a magnetic field.

First, we will discuss 
the asymptotics for small wave vectors of the zero-frequency
density-density response function by a method which allows us to study
the decoupling of correlation functions in a Singwi-Sj\"olander
theory of the half filled lowest Landau level. 
Singwi and Tosi used in \cite{si2} a method to obtain the compressibility sum
rule for the Coulomb system from the Heisenberg equation of motion of the
longitudinal momentum operator $ \vec{q}/q \cdot \vec{\pi}(\vec{q}) $.
The commutator $ \mbox{CH} $ of $ \vec{q}/q \cdot \vec{\pi}(\vec{q}) $  with the
Hamiltonian (\ref{40}) is
\begin{align}
&\mbox{CH}(\vec{q},t) =\sum\limits_{\vec{k}\,'} \frac{1}{m_b}\frac{\left(\vec{k} \vec{q}\right)^2}{q}
 a^+_{\vec{k}+\vec{q}/2}(t) a_{\vec{k}-\vec{q}/2}(t)  \eqlabel{50} \\
&  +\sum\limits_{\vec{q}\,'\not=0}\left(v_0(q')+v_2(q')\right)
 :\rho(\vec{q}\,',t)\rho(\vec{q}-\vec{q}\,',t): \left(
  \frac{\vec{q}\,'\vec{q}}{ q} \right)  \nonumber \\
& +i \sum\limits_{\vec{q}\,'} \frac{v_1(q\,')}{m_b} 
 :\left(\frac{\vec{q}\,'}{q'} \times \vec{\pi}(\vec{q}\,',t)\right)
 \rho(\vec{q}-\vec{q}\,',t):\left(\frac{\vec{q}\,'
     \vec{q}}{q} \right)
                           \nonumber \\
&   -i\sum\limits_{\vec{q}\,'\not=0}  \frac{v_1(q\,')}{m_b} \Bigg[
 : \left(\frac{\vec{q}\,'}{q'} \times \vec{\pi}(\vec{q}+\vec{q}\,',t)\right)
      \rho(-\vec{q}\,',t):
       \left(\frac{\vec{q}\,'\vec{q}}{q} \right) \nonumber    \nonumber  \\
 &+
:  \left(\frac{\vec{q}}{q} \cdot \vec{\pi}(\vec{q}+\vec{q}\,',t)\right)
 \rho(-\vec{q}\,',t)
 :\left(\frac{\vec{q}\,'}{q'}\times \vec{q} \right) \Bigg]\,.
 \nonumber
 \end{align}
We consider in  addition a coupling to 
an external potential $ V^{\mbox{\scriptsize ext}}(\vec{q},t) $.
Since we want to study the adiabatic limit
we neglect the time derivative  
in the Heisenberg equation of motion 
\begin{equation}
\sum\limits_{\vec{q}\,'} V^{\mbox{\footnotesize ext}}(\vec{q},t) \rho(\vec{q}-\vec{q}\,',t)
\left(\frac{\vec{q}\,'\vec{q}}{q} \right)=- \mbox{CH}(\vec{q},t) \,.  \eqlabel{60}
\end{equation}
We take the expectation value of this equation of motion 
with respect to the ground state of the system
(ground state for $ V^{  \mbox{\scriptsize ext}}=0 $)
and get the following equation which is valid to linear order in
$ V^{\mbox{\footnotesize ext}}$ 
\begin{align}
&V^{ \mbox{\footnotesize ext}}(\vec{q},t) \rho_0
=-\langle \mbox{CH}(\vec{q},t)\rangle_c \frac{1}{q}  \eqlabel{70}  \\
&-\rho_0 \left(v_0(q)+v_2(q)\right)
\langle \rho(\vec{q},t) \rangle
-i \rho_0 \frac{v_1(q)}{m_b}
\left(\frac{\vec{q}}{q} \times \langle \vec{\pi}(\vec{q},t) \rangle \right). 
  \nonumber 
\end{align}
$ \langle \cdot \rangle_c$ is the cumulant part of the expectation value
$ \langle AB \rangle_c=\langle AB \rangle-\langle A \rangle \langle B
\rangle$ where $ A $, $ B $ are  the operators $ \vec{\pi}$ or $\rho$.

In the following we will discuss the leading order in $ q $ of the terms in
(\ref{70}).  
By considering only the linear order in $V^{
  \mbox{\footnotesize ext}}$ we get for $ q  \to 0 $  the following
relation (compare \cite{si2}) 
\begin{equation}
\langle O \rangle = \left[\frac{\partial}{\partial \rho_0}
  \langle O \rangle^{
  \mbox{\footnotesize $V^{
  \mbox{\scriptsize ext}}$}=0} \right]\langle \rho(q,t)\rangle \, ,
\eqlabel{80}         
\end{equation}
for an operator $ O $. Here $ \langle \cdot \rangle^{
  \mbox{\footnotesize $V^{
  \mbox{\footnotesize ext}}$}=0} $ is given by the expectation value for
$V^{\mbox{\scriptsize ext}}=0$. We obtain that
$ \langle \mbox{CH}(\vec{q},t) \rangle_c /q $ has the
asymptotic behaviour $ O(q^0) \langle \rho(\vec{q}) \rangle  $ for
$ q \to 0 $. For deriving this asymptotics we take
the $ q \to 0 $ limit in every cumulant expectation value in (\ref{50}). 
Every additive term in (\ref{50})
contains either a term proportional to $ q $ or has a
term linear in $ \vec{q}\,' \cdot \vec{q}/q $ (the terms with a quadratic
$ q' $ in the numerator cancels).  
We now discuss the leading terms in (\ref{70}).
We begin with the third term of the right hand side of equation
  (\ref{70}) and retransform it with the help of (\ref{10})
  from the CS fermions to the electrons (CS retransformation).
  Denoting the expectation value with respect
 to the electronic ground state by $ \langle \cdot \rangle_e $ we obtain 
\begin{align}
& \langle \vec{\pi}(\vec{q},t)\rangle=  \eqlabel{90}\\
& m_b \langle \vec{J}(q,t)\rangle_e-i \sum\limits_{\vec{q}\,' \not=0} \vec{e}_z
   \times \frac{\vec{q}\,'}{{q\,'}^2}\langle : \rho(\vec{q}\,'+\vec{q}) \rho(-\vec{q}\,')
   :\rangle_e \,. \nonumber 
\end{align}
$ J(\vec{q},t)$ is the electron current operator $ J(\vec{q},t)=\sum_{\vec{k}}
(\vec{k}/m_b) a^+_{e,\vec{k}+\vec{q}/2}(t) a_{e,\vec{k}-\vec{q}/2}(t)+
(1/m_b)\sum_{\vec{q}\,'}\vec{A}(\vec{q}\,') \rho(\vec{q}-\vec{q}\,',t) $.
Here $\vec{A}(\vec{q}\,') $ is the external vector potential of the $ \nu=1/2 $
system.
Now we  use (\ref{80}) to calculate the first term of
the right hand side of (\ref{90}).
Since $ \langle \vec{J}(\vec{r},t)\rangle_e^{
  \mbox{\scriptsize  $V^{
  \mbox{\scriptsize  ext}}\nts{3} =0 $}}\nts{1}=0$ for all electron densities of the system
one gets zero for this term. 
The cumulant expectation value of the second term of
(\ref{90}) is of order $ O(q^0) \langle \rho(\vec{q}) \rangle $.
Thus, the third term of (\ref{70}) is given ($ q \to 0 $) by the non-cumulant
part  
\begin{eqnarray}
 i \rho_0 \frac{v_1(q)}{m_b}
\left( \frac{\vec{q}}{q}  \times 
  \langle \vec{\pi}(\vec{q},t) 
  \rangle \right) & = &   
- \rho_0 v_2(q) \langle \rho(\vec{q},t) \rangle  \eqlabel{100} \\
& & +O(q^0)\langle \rho(\vec{q},t)
\rangle \;.  \nonumber 
\end{eqnarray}
By inserting this equation in (\ref{70}) we get for the static linear response
in the limit $ q \to 0 $ 
\begin{equation}  
 \langle \rho(\vec{q},t)\rangle=-\frac{1}{(v_0(q)+O(q^0))}  V^{\mbox{\scriptsize ext}}(\vec{q},t) \;. \eqlabel{110}
\end{equation}
Thus the static density-density response function vanishes as $ v_0(q)^{-1}
 $, i.e linearly for $ q \to 0 $. 

The above results are rigorous consequences of the equation of motion in
the limit $ q \to 0 $. We consider this result as a consistency condition
for an approximative calculation of the correlation functions:
An approximative density-density correlation function and the corresponding
density-momentum correlation function have to satisfy the above analysis.  
Especially we have seen that the $ 1/q^2 $ singularity of the non
cumulant part of the
commutator in (\ref{70}) , due to the commutation with the $ v_2 $-term of $
H $ (second term in (\ref{50})), is cancelled by the $ 1/q^2 $ singularity of the non cumulant part,
due to the commutation with the $ v_1 $-term of $ H $ (third term in
(\ref{50})). Thus there is no $
1/q^2 $-singularity in the denominator of (\ref{110}). 
It is clear from the structure
of the operators and from the derivation above that the first term gets
this $ 1/q^2 $ singularity by averaging with respect to every state not only
the CS ground state.
This is no longer true for the second term. In this term one gets the
$ 1/q^2 $ singularity by averaging over the CS ground state.
This ground state is reached by the dynamics of the CS system.
Thus the consistency relation is written for $ q \to 0 $ as
\begin{align}
&\sum\limits_{\vec{q}\,'\not=0} v_2(q')
\langle  :\rho(\vec{q}\,',t)\rho(\vec{q}-\vec{q}\,',t):\rangle  \left(
  \frac{\vec{q}\,'\vec{q}}{  q} \right)  \nonumber \\
& +i \sum\limits_{\vec{q}\,'} \frac{v_1(q\,')}{m_b} 
 \langle :\left(\frac{\vec{q}\,'}{q'} \times \vec{\pi}(\vec{q}\,',t)\right)
 \rho(\vec{q}-\vec{q}\,',t): \rangle \left(\frac{\vec{q}\,'
     \vec{q}}{q} \right)
                           \nonumber \\
&   -i\sum\limits_{\vec{q}\,'\not=0}  \frac{v_1(q\,')}{m_b}
\langle  : \left(\frac{\vec{q}\,'}{q'} \times \vec{\pi}(\vec{q}+\vec{q}\,',t)\right)
      \rho(-\vec{q}\,',t): \rangle 
       \left(\frac{\vec{q}\,'\vec{q}}{q} \right)  \nonumber \\
&=  O(q) \;. \eqlabel{115} 
      \end{align}
Here, the third term cancels the cumulant part of the second term 
to order $ O(q)$.

We add a remark concerning the Hamiltonian (\ref{40}).
The Hamiltonian (\ref{40}) is truncated by keeping only terms up to second
order in the quadratic density fluctuations.
We can show by the same methods as above
that an analysis of the full problem with the Hamiltonian
(\ref{20}) leads to the same results.

In the next section we formulate a Singwi-Sj\"olander theory of the
half-filled Landau level, where we make an approximation of the
momentum-density and density-density correlation function which respects the
consistency condition (\ref{115}) derived above.

\section{CS response function which includes  dynamical correlations}
In this section, we will calculate response functions of the $ \nu=1/2 $
CS system which include correlations beyond the RPA.
As mentioned earlier, one 
transforms the Hamiltonian of electrons in a magnetic
field to a CS-Hamiltonian (\ref{40}) at zero magnetic field. So, one can
calculate response functions of the $\nu=1/2 $ system with approximative 
methods which
were developed for
the Coulomb system earlier. In this section, we will apply  the theory of
STLS \cite{si1} to the CS system in the spirit of CC \cite{co1}.
The response function matrix $ \chi $  relates the density $ \rho(\vec{k},
\omega) $ and transverse
momentum  response $ \pi_{\mbox{\scriptsize T}}(\vec{k},\omega)=\vec{k}/k \times
\vec{\pi}(\vec{k}, \omega) $ to an external perturbation,
a scalar potential $ V^{\mbox{\footnotesize ext}} $ and a transverse
vector potential  
$ A_{\mbox{\scriptsize T}}^{ \mbox{\footnotesize ext}} $ via
 \begin{equation}  
\left( \begin{array}{c}
    \rho(\vec{k},\omega) \\
    \pi_{\mbox{\scriptsize T}}(\vec{k},\omega)
    \end{array} \right)= \left(\chi(\vec{k},\omega)\right) \cdot
  \left( \begin{array}{c}
  V^{ \mbox{ \scriptsize ext}}(\vec{k},\omega)   \eqlabel{118}  \\
  A_{\mbox{\scriptsize T}}^{\mbox{\scriptsize ext}}(\vec{k},\omega)   
    \end{array} \right) \,.
\end{equation}
Following the original derivation of STLS \cite{si1} we start from
the equation of motion for the one-body Wigner distribution function
\begin{equation}
  f^{(1)}(\vec{r}, \vec{p};t)=\sum\limits_{\vec{k}} e^{i \vec{k} \vec{r}}
  \langle a^+_{\vec{p}-\vec{k}/2}(t) a_{\vec{p}+\vec{k}/2}(t) \rangle 
\eqlabel{120}
\end{equation}
which determines the density $ \rho(\vec{r},t)=\sum_{\vec{p}}
f^{(1)}(\vec{r}, \vec{p};t) $ and the momentum
$ \vec{\pi}(\vec{r},t)=\sum_{\vec{p}} \vec{p}
f^{(1)}(\vec{r}, \vec{p};t) $.
The Heisenberg
equation of motion for $ f^{(1)}(\vec{r}, \vec{p};t)$ is 
\begin{align}
 &\frac{\partial}{\partial t} f^{(1)}(\vec{r}, \vec{p};t)=
\frac{\vec{p} \cdot \vec{\nabla}_{\vec{r}}}{m_b} f^{(1)}(\vec{r}, \vec{p};t)
  \eqlabel{130} \\
 &+ \int d^2r\,' \sum\limits_{\vec{p}\,'} 
 \Big[ \frac{(\vec{p}-\vec{p}\,')_j}{m_b}(\nabla_{r,i} \phi_j)(\vec{r}-\vec{r}\,')
  \nabla_{p,i} \nonumber \\
 &+ \left[ \nabla_{r,i}(v_0+v_2)\right](\vec{r}-\vec{r}\,') \nabla_{p,i}\Big]
 f^{(2)}(\vec{r}, \vec{p}; \vec{r}\,'\nts{2}, \vec{p}\,';t)
 \nonumber \\
 &+ \nabla_{p,i}f^{(1)} \nabla_{r,i}
 V^{\mbox{\footnotesize  ext}}(\vec{r},t)+ \frac{p_j}{m_b}
 \nabla_{p,i}f^{(1)}\nabla_{r,i}
 A_j^{\mbox{\footnotesize ext}}(\vec{r},t) \, ,\nonumber
\end{align}
where $ \nabla_{p,i}= \partial / \partial p_i $, and
\begin{eqnarray}
f^{(2)}(\vec{r}, \vec{p};\vec{r}\,' \nts{2}, \vec{p}\,';t) & = & 
 \sum\limits_{\vec{k}, \vec{k}\,'} e^{i \vec{k}\,' \vec{r}\,'}
 e^{i \vec{k} \vec{r}}   \langle : a^+_{\vec{p}-\vec{k}/2}(t)
 a_{\vec{p}+\vec{k}/2}(t) \nonumber  \\
& &\times \, a^+_{\vec{p}\,'-\vec{k}\,'/2}(t)
 a_{\vec{p}\,'+\vec{k}\,'/2}(t) :\rangle \;. \eqlabel{140}
\end{eqnarray}
is the two-body distribution function. \\
Now we have to decouple
the correlation function
$ f^{(2)}(\vec{r}, \vec{p}; \vec{r}\,'\nts{2}, \vec{p}\,';t) $ in (\ref{130}).
For the Coulomb theory STLS uses the following decoupling
\begin{equation}
f^{(2)}(\vec{r}, \vec{p}; \vec{r}\,'\nts{2}, \vec{p}\,';t) \approx   
f^{(1)}(\vec{r}, \vec{p}; t) f^{(1)}(\vec{r}\,', \vec{p}\,';t)
g(\vec{r}-\vec{r}\,') \, . \eqlabel{150}
\end{equation}
If one uses the decoupling function $ g(\vec{r}-\vec{r}\,')=1 $ in
(\ref{150}) it is easily seen that one gets with the help of
this decoupling the density-density response
function in RPA from the equation of motion (\ref{130}).
Such a decoupling does not respect the small distance correlations.
STLS take into account these correlations by taking for
$ g(\vec{r}-\vec{r}\,') $
the equilibrium, static pair correlation function.
This could also be motivated as follows.
One can determine $ g(\vec{r}-\vec{r}\,')$ by 
summing (\ref{150}) over $ \vec{p}, $ and $\vec{p}\,'$ such that
both sides of the
approximation
(\ref{150}) coincide for
$ V^{\mbox{\scriptsize ext}}=A^{ \mbox{\scriptsize ext}}_{
  \mbox{\scriptsize T}}=0 $ :
\begin{equation}
g(\vec{r}-\vec{r}\,')=\frac{\langle:\rho(\vec{r}) \rho(\vec{r}\,'):\rangle}
{\langle \rho(\vec{r})\rangle \langle \rho(\vec{r}\,')\rangle}\, .   \eqlabel{160}
\end{equation}
The ansatz (\ref{150}), (\ref{160}) is a decoupling, specific for
the Coulomb theory because the interaction of the Coulomb Hamiltonian
consists only of a density-density vertex.
In the CS theory the Hamiltonian (\ref{40}) has  
in addition to the density-density vertex
a density-momentum vertex. The effect of this is given by the
first term in the square brackets on the right hand side of
equation (\ref{130}). It is not
clear whether the decoupling (\ref{150}), (\ref{160}) is a good approximation
for this term. In their approximation
CC \cite{co1} used the decoupling (\ref{150}),
(\ref{160}) for the CS theory.

In the following, we want to check  if this
decoupling respects the consistency condition
(\ref{115}). For this, we have to determine the density-density correlation
function and the density-momentum correlation function from the
decoupled two body Wigner distribution function.  
By a summation over $ \vec{p} $, $\vec{p}\,' $ we get from (\ref{150}) 
\begin{eqnarray}
\langle : \rho(\vec{r}\,',t) \rho(\vec{r},t) : \rangle &\approx  &
\langle   \rho(\vec{r}\,',t)\rangle  \langle \rho(\vec{r},t) \rangle
\, g(\vec{r}\,'-\vec{r}) ,\eqlabel{163} \\ 
\langle :  \vec{\pi}(\vec{r}\,',t) \rho(\vec{r},t) : \rangle &\approx  &
\langle \vec{\pi}(\vec{r}\,',t)\rangle  \langle \rho(\vec{r},t) \rangle
\, g(\vec{r}\,'-\vec{r}) .\eqlabel{165}
\end{eqnarray}
By inserting these decouplings into (\ref{115}) we get for the first summand
$  \rho_0 v_2(q) q \langle \rho(\vec{q},t) \rangle +O(q)$.
The third summand is zero. The second summand of (\ref{115}) is given by
a similar discussion as in the last section (see especially the discussion
below equation (\ref{90}))
\begin{align}
&i \rho_0 \nts{1}\sum\limits_{\vec{k}}  \frac{v_1(\vec{q}+\vec{k})}{m_b}
\nts{1} \left(\frac{(\vec{q}+\vec{k})}{|\vec{q}+\vec{k}|} \nts{1}\times \nts{1}
  \langle \vec{\pi}(\vec{q},t) \rangle \right) \nts{1}
(S(k)-1)  \frac{ (\vec{q}+\vec{k}) \vec{q}}{q} 
 \nonumber \\
 &-\rho_0 v_2(q) q  \langle  \rho(\vec{q},t) \rangle
+O(q)\,.  \eqlabel{167}
\end{align}
where $ S(k)-1 $ is the Fourier transformion of $ g(r)-1 $. 
With the help of $ \sum_{\vec{k}}( S(k)-1) =-1 $ we get for 
(\ref{167}) $\, - (1/2)\rho_0 v_2(q) q \langle \rho(\vec{q},t) \rangle
+O(q)$. Thus we obtain that the decoupling
of CC does not fulfill the consistency condition (\ref{115}).  
That is the reason why CC obtained in the scheme of their decoupling
a zero-frequency
density-density response function which vanishes as the
square of the wave vector in the long wavelength limit.

To get a better insight into the violation of the consistency
condition (\ref{115}), we calculate the Fourier transformation
of the first two terms in (\ref{115}) multiplied by $q$. 
This is given by 
\begin{align}
&-\int \nts{3} d^2r'
 \nabla_{r,i} \nts{1}\left\{\langle : \nts{2}(\rho(\vec{r}\,',t)-\rho_0) \,
 \rho(\vec{r},t) \nts{2} : \rangle
\nabla_{r,i} v_2(\vec{r}\,'-\vec{r})\right\} \eqlabel{168}  \\
& -\int \nts{3} d^2r'  \nabla_{r,i} \nts{1} 
\left\{ \langle : \nts{2} (\vec{\nabla}_{r'} \times \vec{\pi}(\vec{r}\,',t))
 \, \rho(\vec{r},t) \nts{2} : \rangle 
 \frac{\nabla_{r,i} v_2(\vec{r}\,'-\vec{r})}{4 \pi \rho_0}
 \right\}. \nonumber 
\end{align}
Because of the appearance of the $\vec{\pi} $ in the second term in
(\ref{168})
we proceed similar as in the discussion below (\ref{90})
by retransforming to the electronic system.
Then, the term of quadratic
density fluctuations cancels the first term in (\ref{168}).
This is no longer true by using the decouplings (\ref{163}), (\ref{165})
in (\ref{168}). The operator  $ \vec{\nabla}_{r'}  $ in the second term in
(\ref{168}) acts in this decoupling  on $ \langle \pi(\vec{r}\,')\rangle
\langle \rho(\vec{r}) \rangle$ as
well as on $ g(\vec{r}\,'-\vec{r}) $. This results in two summands.
When we take  
the quadratic density fluctuation part after a CS retransformation
of these two terms we get that the first term is cancelled by the   
first term in (\ref{168})
(with the help of  $ \vec{\nabla}_{r'} \times \vec{\phi}(\vec{r}\,'-\vec{r})=2
\delta(\vec{r}\,'-\vec{r})$). The term with $ \vec{\nabla}_{r'}  $
acting on $ g(\vec{r}\,'-\vec{r}) $
is the reason for the violation of the consistency condition.
In other word doing the CS-retransformation first and then the decoupling
of the correlation function, or vice versa, leads to different results.
We require that these two actions commute and that fixes
the density-momentum decoupling in the case of a given decoupling of the
density-density correlation function. For the given density-density
decoupling
of STLS  (\ref{163}) we get by this requirement the decoupling
\begin{equation}
\langle : \nts{1} \vec{\nabla}_{r'} \times \pi(\vec{r}\,',t) \rho(\vec{r},t)
\nts{1}: \rangle
\nts{1}\approx \nts{1}  
g(\vec{r}\,'-\vec{r})  \langle (\vec{\nabla}_{r'} \times
\vec{\pi}(\vec{r}\,',t))
\rangle  \langle \rho(\vec{r},t) \rangle .\eqlabel{180}
\end{equation}
Thus we get that 
$g(\vec{r}\,'-\vec{r})$
should not be differentiated. 
In the decoupling of the equation of motion (\ref{130}), we get for the
second  term 
in the square brackets after a Fourier transformation
with respect to $ \vec{r} $:  
\begin{align}
&\int d^2r \,d^2r' \,\sum\limits_{\vec{p}\,'}\,  e^{i \vec{q} \vec{r}}
[\nabla_{r,i} v(\vec{r}-\vec{r}\,')] \nabla_{p,i}
 f^{(2)}(\vec{r}, \vec{p}; \vec{r}\,', \vec{p}\,';t) \nonumber \\
& \approx \int d^2r \, d^2r'\sum\limits_{\vec{p}\,'} e^{i \vec{q} \vec{r}}
[\nabla_{r,i} v(\vec{r}-\vec{r}\,')]   \nonumber \\
& \qquad \cdot  \nabla_{p,i}f^{(1)}(\vec{r}, \vec{p};t) f^{1}(\vec{r}\,', \vec{p}\,';t)
 \,g (\vec{r}-\vec{r}\,')  \eqlabel{170}
 \end{align}
where $ v=v_0+v_2 $.
The first term in the square brackets should be decoupled by 
\begin{align}
&\int d^2r \,d^2r' \sum\limits_{\vec{p}\,'} \,e^{i \vec{q} \vec{r}}
  \frac{(\vec{p}-\vec{p}\,')_j}{m_b}(\nabla_{r,i} \phi_j)(\vec{r}-\vec{r}\,')
\nonumber \\
& \qquad \cdot   \nabla_{\vec{p},i}f^{(2)}(\vec{r}, \vec{p}; \vec{r}\,'
\nts{2}, \vec{p}\,';t)
 \eqlabel{190}  \\
&=   \int d^2r \,d^2r' \sum\limits_{\vec{p}\,'} \,e^{i \vec{q} \vec{r}}
(\nabla_{r,i} v_2(\vec{r}-\vec{r}\,')) \frac{1}{4 \pi \rho_0 }
\nonumber \\
&\cdot \left(\vec{\nabla}_r \times \vec{p}+ \vec{\nabla}_{r'} \times \vec{p}\,'
+i \vec{q} \times \vec{p} \right) \nabla_{\vec{p},i}
f^{(2)}(\vec{r}, \vec{p}; \vec{r}\,' \nts{2}, \vec{p}\,';t) \nonumber \\
&\nts{1} \approx   \nts{1} \int d^2r \,d^2r' \sum\limits_{\vec{p}\,'} \,e^{i \vec{q} \vec{r}}
(\nabla_{r,i} v_2(\vec{r}-\vec{r}\,')) \frac{1}{4 \pi \rho_0 }
g (\vec{r}-\vec{r}\,')
\nonumber \\
&\cdot \nts{2} \left(\vec{\nabla}_r \nts{1}\times \nts{1}\vec{p}+ \vec{\nabla}_{r'}\nts{1} \times \nts{1}\vec{p}\,'
+i \vec{q}\nts{1} \times \nts{1} \vec{p} \right)\nts{2} \nabla_{\vec{p},i}
f^{(1)}(\vec{r}, \vec{p};t) f^{(1)}(\vec{r}\,', \vec{p}\,';t),
  \nonumber 
\end{align}
where $ g(\vec{r}-\vec{r}\,') $ is not differentiated.

With the help of the Heisenberg equation of motion (\ref{130}) and  the
approximations (\ref{170}) and (\ref{190}) it is possible to calculate the
response matrix (\ref{118}) (similar to the calculations of CC \cite{co1}).
By doing this we get
\begin{equation}
\chi= \chi^0\left[1-U \chi^0\right]^{-1},   \eqlabel{210}
\end{equation}
where
\begin{equation}           \eqlabel{220}
\chi^0=\left(
  \begin{array} {c c}
    \chi^0_{\rho \rho} & 0 \\
    0 & \chi^0_T
  \end{array} \right)\,.
\end{equation}
$ \chi^0_{\rho \rho} $ is the ideal gas density-density response and
$ \chi^0_T $ is the corresponding transversal momentum-momentum response.
These response functions are known analytically
\cite{di1}. The matrix of the effective potentials is
\begin{equation}   \eqlabel{230}
   U= \left(\begin{array} {c c}
       w_0(k)+w_2(k) & i w_1(k) \\
       -i w_1(k) & 0
       \end{array}\right)\;, 
\end{equation}       
 where $ w_{\alpha}(k)=(1-G_{\alpha}(k)) v_{\alpha}(k) $ and the local field
 factors $ G_{\alpha}(k) $ are given by
\begin{equation}  
  G_{\alpha}(k)=\frac{1}{\rho_0}\sum\limits_{\vec{p}} \left(1- S(p)\right)
   \frac{[ \vec{k} \cdot
      (\vec{k}-\vec{p}) ]^{(a_\alpha+b_\alpha)/2}}{
    k^{a_\alpha} |\vec{k}-\vec{p}|^{b_\alpha}}. \eqlabel{240}
 \end{equation}
Here  
 $ a_0=b_0=1$,
 $ a_1=0 $,
 $ b_1=2 $ and
 $ a_2=0 $, $ b_2=2 $.
 
In the following, we denote the quantities which are calculated by the
 means of CC with an upper index  CC, for comparison.
The structure factor $ S(k) $ can be calculated
with the help of the fluctuation-dissipation theorem \cite{mah1}
 \begin{equation} \eqlabel{250}
   S(k)=-\frac{1}{ \rho_0 \pi} \int_0^{\infty} d\omega \; \mbox{Im}
   [\chi_{\rho \rho}(k,\omega)] \, ,
 \end{equation}
 where $ \chi_{\rho \rho} $ is density-density part of the response matrix
 $ \chi $ in (\ref{118})
\begin{equation}
\chi_{\rho \rho}(k,\omega)=\frac{\chi^0_{\rho \rho}(k,\omega)}{
  1-\chi^0_{\rho \rho}(w_0(k)+w_2(k)+w_1(k)^2 \chi_T^0)} \,.\eqlabel{260}
\end{equation}
By using  their decoupling, CC calculated a similar response as in
 (\ref{230}) and (\ref{240}) with
 $ a_0^{\mbox{\scriptsize CC} }=b_0^{\mbox{\scriptsize CC}}=1$,
 $ a^{\mbox{\scriptsize CC}}_1=1 $, $ b^{\mbox{\scriptsize CC}}_1=1 $ and
 $ a^{\mbox{\scriptsize CC}}_2=0 $, $ b^{\mbox{\scriptsize CC}}_2=2 $.
 Using the rotational invariance of $ S(k) $ it is easy to show that
 $ G_0(k)=G^{\mbox{\scriptsize CC}}_0(k)$
 is linear in $ k$. Further $  G_1(k)=
G^{\mbox{\scriptsize CC}}_2(k)=G_2(k) $ is
quadratic in $k $ and $ G^{\mbox{\scriptsize CC}}_1(0)=1/2 $. Thus we obtain
for our decoupling a zero-frequency
density-density response function which vanishes linearly in the
long wavelength limit. For the CC decoupling it
 vanishes as the square of the wave vector in the long wavelength limit.
 Since the only difference
 in the $ G $-terms of our decoupling and the decoupling of CC is given by
 $ G_1 $, we will compare $ G_1 $ for these two decouplings.
 One gets from the integral (\ref{240}) that
 $ G_1(k) $
 approaches to $ G^{\mbox{\scriptsize CC}}_1(k) $ for 
 $k \gtrsim k_F$.
 Both functions are equal for  $ k \to \infty $. As mentioned above
 this is not the case for $ k\to 0 $. 
 
\begin{figure}[b]
  \psfrag{SI2}{}
  \psfrag{SI1}{\scriptsize   CC  }
  \psfrag{RPA}{\scriptsize   RPA  }
 \centerline{\psfrag{xxx}
   {\smash{\raisebox{-0.05cm}{\footnotesize $\nts{10}(\omega 2m)/k_F^2 $}} }
   \psfrag{y}{\turnbox{180}{\footnotesize  $\nts{15} \pi \rho_0\,  S(k,\omega)$ }}  
 \epsfig{file=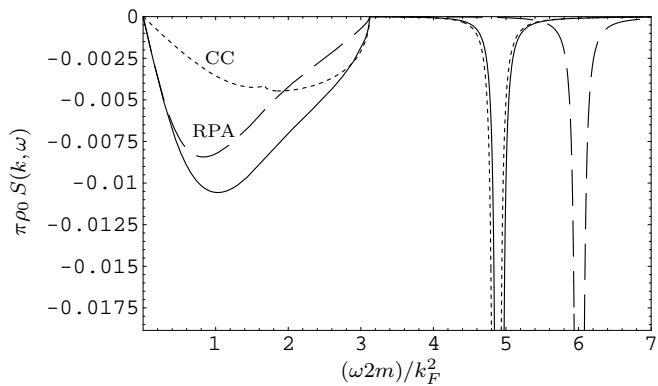,width=8.5cm}
  }
\hspace*{0.1cm}
 \caption{The dynamic structure factor $ S(k,\omega) $
   times $ \pi \rho_0 $ for
   $k=0.6k_F$ and $ r_S=6 $, as a function of $ \omega 2m/k_F^2 $ for
   RPA, CC and our decoupling (solid line).
   The $ \delta $-function peak corresponding to the inter Landau level mode
   has been artificially broadened for clarity and contains most of the
   spectral strength.
   \label{fig1}}
 \end{figure}

In the following, we will  calculate a solution of the (integral) equations
 (\ref{240}-\ref{260}) by a numerical
 iteration method.
 As a starting point we use $ G_{\alpha}=0 $ for all $\alpha$,
 corresponding to the RPA.
For doing this one has to choose the dimensionless coupling strength
$ r_s=1/(a_0 \sqrt{\pi \rho_0}) $, where $ a_0 $ is the
Bohr radius $ a_0=\epsilon^2 m_b$. Since the results show little
 variation with this coupling strength (in the region
 $ 1 \lesssim r_s \le 10 $)
we will choose for definiteness $ r_s=6 $ in the following
figures. 
 This special choice of $ r_s $ could be motivated through calculations of
 the effective mass of the $ \nu=1/2 $ system \cite{co1}.
In the following we present and discuss our results for various quantities
 derived from the density-density response function $ \chi_{\rho \rho} $.
In Fig. \ref{fig1} we show $ (\pi \rho_0 ) S(k,\omega) $, where
\begin{equation}\eqlabel{270}
S(k,\omega)=-\frac{1}{\pi \rho_0 } \mbox{Im}\,[\chi_{\rho \rho}(k, \omega)]
\end{equation}
is the dynamical structure factor.
With the help of
$ \chi^0_{\rho \rho}= \rho_0 k^2/(m_b \omega^2)$ and
$ \chi_{T}^0= 2 \pi\rho_0^2 k^2 /(m_b^3 \omega^2) $ which is valid
for $ k/k_F \ll 1$ ,
$ kk_F/m_b \ll \omega $, we get from (\ref{260}) that $ S(k,\omega)$ 
has a pole at the cyclotron
frequency $ \omega=k_F^2/m_b $ describing the inter Landau
level excitation,
which is unaffected  at $ k=0 $ by correlations and is in agreement with 
Kohn's theorem.
For finite values of $ k $ we get from Fig.
\ref{fig2} that the cyclotron mode (the pole of $ S(k,\omega) $) of CC and
our decoupling 
is given by a smaller frequency than in the RPA. 
From Fig. \ref{fig1} we obtain that
$ S(k, \omega) \approx S^{\mbox{\scriptsize CC}}(k, \omega) $
for $ \omega \gg  kk_F/m_b  $.
\begin{figure}[t]
  TABLE I. The Coulomb interaction  energy
  $ \langle e^{\mbox{\scriptsize Coul}}(r_s)\rangle $ per particle \\[0.1cm]  
\begin{tabular}{c | c c  c c }
  $ r_s $ &\mbox{This Work}   &CC & RPA   \\ \hline 
  $ \to 0 $   &  $  -1.00 \epsilon^2/(a_0 r_s) $ &$  -1.00 \,\epsilon^2/
  (a_0 r_s) $
  &  $ -1.19 \,\epsilon^2/(a_0 r_s) $ \\
  $  6 $ & $ -1.04 \, \epsilon^2/(a_0 r_s) $ & $ -1.04 \, \epsilon^2/(a_0
  r_s)$   & $ -1.76 \,\epsilon^2/(a_0 r_s) $  \\ \hline
\end{tabular}
\end{figure}
This can be
understood
by the asymptotic form of  $ \chi^0_{\rho \rho} $ and $ \chi_T^0 $
in this range
and the similarity of $ G_0(k) $ and $ G_2(k) $ for CC and our decoupling for $ k \ll
k_F $. It is also seen from Fig. \ref{fig1} that
$ S(k, \omega) \approx S^{\mbox{\scriptsize RPA}}
(k, \omega) $
for small values of $ \omega 2 m/k_F^2 $.
This can be also seen in Fig. \ref{fig3}, where we show the
function $ \chi_{\rho \rho}(k,0) $. One sees from this figure that 
$ \chi_{\rho \rho}(k, 0) \approx \chi_{\rho \rho}^{\mbox{\scriptsize RPA}}(k,0) $
for $ k \lesssim k_F$ and
$ \chi_{\rho \rho}(k, 0) \approx \chi_{\rho \rho}^{ 
  \mbox{\scriptsize CC}}(k,0) $ for $ k \gtrsim k_F$.
This is understandable by the asymptotic forms of the $ G(k) $s
(see the discussion above).
One also obtains from this figure that $ \chi^{\mbox{\scriptsize CC}}(k,0)$
has the asymptotic $ O(k^2) $ for $ k \to 0 $.
\begin{figure}[b]
 \psfrag{SI2}{}
  \psfrag{SI1}{\scriptsize  CC  }
  \psfrag{RPA}{\scriptsize  RPA  }
\centerline{\psfrag{xxx}
   {\smash{\raisebox{-0.05cm}{\footnotesize $k/k_F$}} }
   \psfrag{y}{\footnotesize \turnbox{180}{
   $\nts {10}(\omega 2m)/k_F^2 $}}   
 \epsfig{file=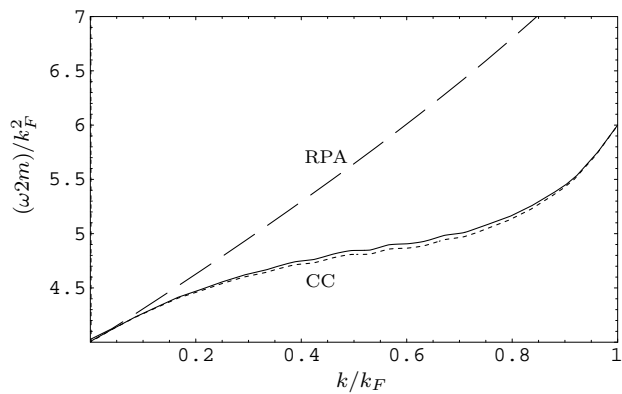,width=8cm}
  }
\hspace*{0.1cm}
 \caption{The cyclotron peak of RPA, CC and our decoupling (solid line)
   for $r_s=6 $.
   It is determined through the
   singularity of Im[$\chi_{\rho \rho}$].\label{fig2}}
   \end{figure}
\begin{figure}[h]
 \psfrag{SI2}{}
  \psfrag{SI1}{\scriptsize  CC  }
  \psfrag{RPA}{\scriptsize  RPA  }
  \centerline{\psfrag{xxx}
   {\smash{\raisebox{-0.05cm}{\footnotesize $k/k_F $}} }
   \psfrag{y}{\turnbox{180}{\footnotesize $ \nts{10} \chi_{\rho \rho}(k,0)$}}  
 \epsfig{file=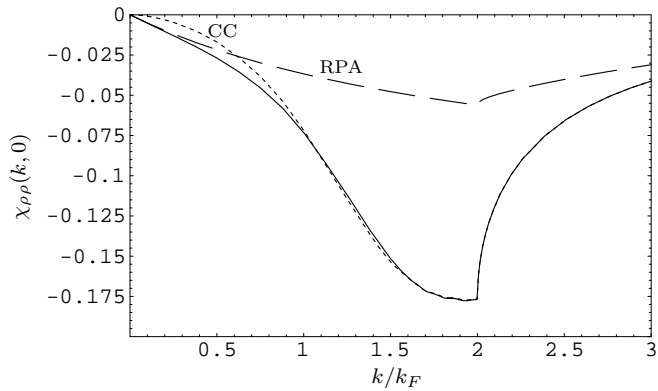,width=8.5cm}
  }
\hspace*{0.1cm}
 \caption{Static density-density response function $ \chi_{\rho \rho}(k,0)$
   as a function of $k/k_F $ for the RPA, CC and our (solid line) decoupling
   scheme for $ r_s=6$.
   The discontinuity in the derivative at $ k=2 k_F $ is due to
   the Fermi surface. \label{fig3}}
\end{figure}

\begin{figure}[b]
 \psfrag{SI2}{}
  \psfrag{SI1}{\scriptsize  CC  }
  \psfrag{RPA}{\scriptsize  RPA  }
\centerline{\psfrag{xxx}
   {\smash{\raisebox{-0.05cm}{\footnotesize $k/k_F $}} }
   \psfrag{y}{\turnbox{180}{\footnotesize $ S(k)$}}  
 \epsfig{file=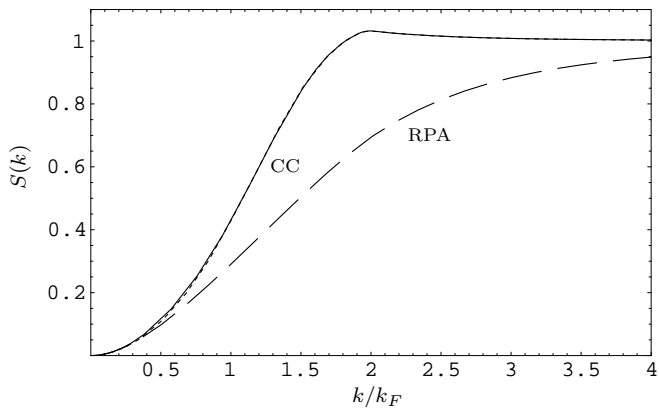,width=8.5cm}
  }
\hspace*{0.1cm}
 \caption{Static structure factor $ S(k) $ as a function of $ k/k_F $
   in RPA, CC and our (solid line)  scheme for $ r_s=6$. \label{fig4}}
\end{figure}

By using the compressibility sum rule \cite{no1,mah1}
\begin{equation}
 K=-\frac{1}{\rho_0^2}\lim_{k \to 0}
 \left(\frac{ \chi_{\rho \rho}(k,0)}
   {1+2 \pi \frac{\epsilon^2}{k} \chi_{\rho \rho}(k,0)}
 \right) \eqlabel{275} 
\end{equation}
we can calculate the compressibility $ K $ of the $ \nu=1/2 $ system.
The denominator is due to the fact that by changing the area of the
system the positive background has to be changed also to conserve
neutrality.   
With the help of $ \chi^0_{\rho \rho}(k,0)=-m_b/(2\pi) $, $
\chi^0_T(k,0)=-\rho_0/m_b $ and (\ref{240}) this leads to   
\begin{equation} 
K=\frac{m_b}{2 \pi \rho_0^2}  
\frac{1}{\frac{16}{3}+
  \frac{m_b}{2\pi \rho_0}
    \langle e^{\mbox{\scriptsize Coul}}(r_s)\rangle } \,.
  \eqlabel{280} 
\end{equation}
We call $  \langle e^{\mbox{\scriptsize Coul}}(r_s)\rangle=1/2 \sum_{\vec{k}}
v_0(k)(S(k)-1) $
the Coulomb interaction  energy per particle, following \cite{mah1}.
For a comparison of the compressibility with the other
  theories we mention that $ K^{\mbox{\scriptsize RPA}}=(3/4)
  (m_b/2\pi \rho_0^2) $ for the RPA and
  $  K^{\mbox{\scriptsize CC}}=0 $ for CC. 
Now we calculate $ \langle e^{
  \mbox{\scriptsize Coul}}(r_s)
\rangle $ for RPA, CC and our decoupling.
The Coulomb interaction energies for $ r_s \to 0 $ and
$ r_s=6 $ are shown in Table I.
We see  from this table that the Coulomb interaction energies of our theory 
and CC are equal.
 By calculating the full Coulomb energy per particle
via coupling
constant integration $ u^{\mbox{\scriptsize Coul}} (r_s)= 1/(a_0 r_s^2)
\int_0^{r_s} dr_s' (a_0 r_s') \langle e^{\mbox{\scriptsize Coul}}(r_s')\rangle $ we get for 
CC and our theory
$ u^{\mbox{\scriptsize Coul}}(r_s)  \approx -1.02 \,\epsilon^2/(a_0 r_s) $.
The Coulomb energy of electrons in the lowest Landau level was calculated
earlier by Morf and d'Ambrumenil \cite{mo1}  and by Girlich \cite{gi1}
by numerical diagonalization methods. Within this method the Coulomb
energy per electron is given by $
u^{\mbox{\scriptsize Coul}}_{\mbox{\scriptsize num}} (r_s) \approx -0.88
\epsilon^2/(a_0 r_s) $.  
This energy has to be compared with
the $ r_s \to 0 $ Coulomb energy of RPA, CC and our method.
We obtain that the $ r_s \to 0 $ energy of the STLS-type methods 
is in a better agreement with the numerical results than the
Coulomb energy of the RPA. (The formula for the coupling constant integration
leads for small $ r_s $ to the equality of the Coulomb interaction energy
with the Coulomb energy.) 
Furthermore, we see from table I
that in the STLS-type theories the Coulomb energy of
the lowest Landau level is a very good approximation to the total Coulomb
energy including higher Landau levels.
This is not the case for the RPA.
To get the reason for the equality of the Coloumb energies of our method and
CC 
we show in Fig. \ref{fig4} for $ r_s=6 $ that $ S(k) =
S^{\mbox{\scriptsize CC}}(k) $ for almost all $ k/k_F $.
We verify that 
all curves in the figure obey the leading $ (k/k_F)^2/2 $
behaviour for small $ k/k_F $,
required by the Kohn theorem. Furthermore, we find for our method that
the $\omega$-momenta
of the dynamical structure factor $ S(k, \omega) $ for small $ k $ (excluding
the cyclotron contribution) coincide with that of the RPA \cite{hlr,mu1}
because for small $k $ our effective potentials coincide with that of the RPA.

At last we calculate the Landau energy.
As in the case of the calculation of the Landau energy in RPA
\cite{di1}, one obtains with the help of a coupling constant
integration for
the Landau energy in our decoupling scheme an infrared diverging energy.
This  is caused by the simplification in considering only
quadratic density fluctuations in the derivation of the Hamiltonian
(\ref{40}) from the 
exact CS Hamiltonian (\ref{20}) (compare the remark at the end of section
II).
The Hamiltonian (\ref{40}) has two defects: (i)
The restriction to the quadratic density fluctuations
leads to an ultraviolet divergence.
(ii) The order of the operators in the product of the two $ \rho$ is changed to
normal order which eliminates the ultraviolet divergence but
results in an infrared divergence \cite{ap1}.
A solution of this problem for the energy in the RPA was given in \cite{di2}
by considering
for the energy calculation all maximal divergent diagrams (RPA) 
together with all first order diagrams of the Hamiltonian (\ref{20}).
The first order 
energy (per particle) of the CS Hamiltonian (\ref{20})
is \cite{sit1}
$ \mbox{uf}^{\mbox{\scriptsize mag}}=4.00 /(m_b (a_0 r_s)^2) $.
In the following, we make a similar calculation for
our decoupling scheme. Thus, we have to
calculate the Landau energy
per particle $  \mbox{u}^{\mbox{\scriptsize mag}}= \mbox{uf}^{\mbox{\scriptsize mag}}+\mbox{ud}^{\mbox{\scriptsize mag}} $
of the Hamiltonian
(\ref{40}) but to extent the first order terms to those of the Hamiltonian
(\ref{20}) with the full three particle interaction.
To this end
we have to multiply  $ v_1(k) $ and $ v_2(k)$
in (\ref{230}) by  a parameter $ \lambda $.
Then the solution for $ \chi $ of the 
(integral) equations (\ref{240}-\ref{260}) depends on $ \lambda $.
From the Hamiltonian  (\ref{40}) we get that for determining
$ \mbox{ud}^{\mbox{\scriptsize mag}} $ one has to calculate
two terms (one contains the density-density response function
$ \chi_{\rho \rho} (k,\omega;\lambda) $, the other contains the
density-momentum reponse function $ \chi_{\rho \pi} (k,\omega;\lambda) $).   
After some algebra one gets for $\mbox{ud}^{\mbox{\scriptsize mag}} $
  through coupling constant integration  
\begin{align}
&\mbox{ud}^{\mbox{\scriptsize mag}}  = \nonumber \\
&- \nts{1}\frac{1}{\rho_0 \pi} \nts{1}
\int\limits_0^1 d\lambda
\sum\limits_{\vec{k}} \int\limits_0^{\infty} \nts{1}
d \omega  \Bigg[\frac{1}{2} v_2(k) \, \mbox{Im}[\chi_{\rho \rho} (k,\omega;\lambda)-
\chi^0_{\rho \rho}(k,\omega)]
  \nonumber  \\
&+v_1(k) w_1(k)\, 
  \mbox{Im}[\chi_{T}^0(k,\omega) \chi_{\rho \rho}(k,\omega;\lambda)] \Bigg]
  \,. 
  \label{290} 
\end{align}
With the help of a numerical integration of (\ref{290}) we get
$\mbox{ud}^{\mbox{\scriptsize mag}}=-2.16 /(m_b (a_0 r_s)^2) $.
The total Landau energy per particle
$ u^{\mbox{\scriptsize mag}} $ is then given by
\begin{equation}\label{300}
 u^{\mbox{\scriptsize mag}}=
 \mbox{uf}^{\mbox{\scriptsize mag}}+
\mbox{ud}^{\mbox{\scriptsize mag}}=1.84\frac{1}{m_b (a_0 r_s)^2} \,. 
\end{equation}   
By doing a similar calculation for the total Landau energy per
particle within the
RPA \cite{di2} we get
$ u_{\mbox{\scriptsize RPA}}^{\mbox{\scriptsize mag}}=1.60/(m_b (a_0 r_s)^2) $.
The Landau energy can be compared to the exact Landau energy
$ \omega_c/2 $  of the $ \nu=1/2 $ system.
This is given by
$ u_{\mbox{\scriptsize ex}}^{\mbox{\scriptsize mag}}=2.00/(m_b (a_0 r_s)^2) $. Thus we obtain that,
in comparison with the RPA, the Landau energy and the Coulomb energy
of the STLS-type theories are in better agreement with the exact Landau
energy and the Coulomb energy (from numerical diagonalization).

\section{Conclusion}
In this paper, we consider an approximation of the response function
of the $ \nu=1/2 $ system which goes beyond the RPA. The method we used
is the STLS-theory, first established for the Coulomb system \cite{si1}.
Recently this theory was applied to the CS system by CC \cite{co1}.
In their theory, CC obtain a density-density response function which vanishes
as the square of the wave vector in the long wavelength limit.
We show in this paper
that the zero-frequency density-density response function vanishes linearly
in the long wavelength limit independent of any approximation.
We obtain this result by considering the
Heisenberg equation of motion for the longitudinal momentum operator.
From this equation of motion, we  derive a consistency condition for a
decoupling of the equal time density-density and
density-momentum correlation functions. We show that this consistency
condition is not fulfilled in the theory of CC and that is the reason for the
quadratic behaviour of
the zero-frequency long wavelength limit of the density-density response.
Based on the functional form of the Heisenberg equation of
motion of the Wigner distribution function (with external potentials),
we suggest a decoupling of the
correlation functions in this equation which respects the consistency
condition.

We solve the decoupled Heisenberg equation of motion
by numerical iterations and get 
the response functions of the theory.
In contrast to the theory of CC, we obtain a
density-density response function which vanishes linearly 
in the long wavelength limit for zero-frequency.
We get agreement for the density-density response  function with the
theory of CC for momenta $ k \gtrsim  k_F $. For $ k \lesssim k_F $ we get 
agreement with the density-density response function  of CC for frequencies
$ \omega \gtrsim  k k_F/m_b $ and to the RPA for
$ \omega  \lesssim  k k_F/m_b $.
Further, we calculate  the compressibility of the theory by using the
compressibility sum rule. We obtain a Coulomb correction to the
compressibility not contained in the RPA.  
With the help of the  response functions, we
calculate the static structure factor, the excitation spectrum and
the Landau as well as the Coulomb energies.
As in the theory of CC, we get density excitations which are lower in their
frequencies as a function of the wave vector than the excitations calculated
with the help of the RPA. The obtained excitation spectrum is almost
identical to the spectrum of CC. 
The same holds for the static
structure factor. This is the reason for the agreement of the lowest
Landau level
Coulomb energy as well as the full Coulomb energy with the energies
calculated by CC. The lowest Landau level Coulomb energy fits better
to the Coulomb
energy calculated by numerical methods \cite{mo1,gi1}
($ \approx 114 \% $ of the numerical lowest Landau level Coulomb energy)
than the RPA.
We remark that the relative part of the Coulomb energy
in the lowest Landau level, i.e. linear in $ \epsilon^2 $,
is much enhanced as compared to the RPA.
Finally, we calculate the Landau energy of the system. We obtain 
a much better approximation of the exact Landau energy (
$ \approx 92 \% $ of the exact Landau energy of the $\nu=1/2 $ system)
than the RPA.

In summary, a consistent decoupling of the Wigner
function in the Heisenberg equation of motion
leads to results  which are in a better agreement with
known numerical and exact results, respectively, than the  RPA.     

\bigskip

We would like to thank S. Conti, M. Hellmund, and W. Apel for many helpful 
discussions during the course of this work. Further we acknowledge
the support of the Graduiertenkolleg ``Quantenfeldtheorie'' at the University
of Leipzig and the DFG Schwerpunktprogramm ``Quanten-Hall-Systeme''.

\end{document}